\documentstyle[epsfig]{mn}

\newcommand{\reference}{\bibitem}

\title[]{Gamma-ray bursts: optical afterglows in the deep Newtonian phase} 

\author[]{Y. F. Huang$^{1,2,3 \; \star}$ and K. S. Cheng$^{2}$
\thanks{E-mail: hyf@nju.edu.cn(YFH); hrspksc@hkucc.hku.hk(KSC) } \\
$^1${\sl Department of Astronomy, Nanjing University, Nanjing 210093, 
         China} \\
$^2${\sl Department of Physics, The University of Hong Kong, 
         Pokfulam Road, Hong Kong, China} \\
$^3${\sl LCRHEA, Institute for High-Energy Physics, Chinese Academy of 
         Sciences, Beijing 100039, China} } 

\date{MNRAS in press (originally submitted in October, 2002) }

\pubyear{2003}

\begin{document}
\voffset=-0.5 in

\maketitle
\begin{abstract}
Gamma-ray burst remnants become trans-relativistic typically in days to 
tens of days, and they enter the deep Newtonian phase in tens of days 
to months, during which the majority of shock-accelerated electrons 
will no longer be highly relativistic. However, a small portion
of electrons are still accelerated to ultra-relativistic speeds and 
capable of emitting synchrotron radiation. The distribution 
function for electrons is re-derived here so that synchrotron 
emission from these relativistic electrons can be calculated. Based 
on the revised model, optical afterglows from both isotropic fireballs
and highly collimated jets are studied numerically, and compared to 
analytical results. In the beamed cases, it is found that, in 
addition to the steepening due to the edge effect and the 
lateral expansion effect, the light curves 
are universally characterized by a flattening during the deep 
Newtonian phase. 
\end{abstract}


\section {Introduction}

Observations of X-ray, optical and radio afterglows from gamma-ray
bursts (GRBs) definitely show that at least most long GRBs are of 
cosmological origin (e.g. Costa et al. 1997; Frail et al. 1997;  
Galama et al. 1997; Metzger et al. 1997; Kulkarni et al. 1998; 
Garcia et al. 1998; Piro et al. 1998; Pedersen et al. 1998; 
Galama et al. 1998a,b; Akerlof et al. 1999; Vreeswijk et al. 1999;  
Zhu et al. 1999; Hjorth et al. 2002). 
The famous fireball model, which incorporates 
internal shocks to account for the main bursts, and external shocks
to account for afterglows, becomes the most popular model  
(M\'esz\'aros \& Rees 1992; Rees \& 
M\'esz\'aros 1992; M\'esz\'aros, Laguna \& Rees 1993; M\'esz\'aros, 
Rees \& Papathanassiou 1994; Katz 1994; Rees \& M\'esz\'aros 1994; 
Sari, Narayan \& Piran 1996). Observed features of GRB afterglows 
can basically be well explained in this frame-work (M\'esz\'aros \& 
Rees 1997; Vietri 1997; Waxman 1997a,b; Wijers, Rees \& M\'esz\'aros
1997; M\'esz\'aros, Rees \& Wijers 1998; Sari, Piran \& Narayan 1998; 
Dai \& Lu 1998, 1999; Dermer, B\"ottcher \& Chiang 1999a; 
Dermer, Chiang \& B\"ottcher 1999b; Wijers \& Galama 1999; 
Wijers et al. 1999). 
However, we are still far from resolving the puzzle of GRBs, since
many crucial information concerning the true nature of 
the inner engine is lost after an initial acceleration phase of 
the fireball evolution. Studies on afterglows can hopefully provide
important clues of GRB progenitors. For example, from afterglow 
observations, we can measure the beaming angle, the environment 
density, the total kinetic energy etc (Hjorth et al. 2002; 
K\"onigl \& Granot 2002). For recent reviews on GRBs,
see Piran (1999), van Paradijs, Kouveliotou \& Wijers (2000) and
M\'esz\'aros (2002).

As the most violent bursts in the Universe since the Big Bang, 
GRBs are most impressing for their extremely relativistic motions, 
with Lorentz factor $\gamma \geq$ 100 --- 1000. In 1997, Wijers  
et al. (1997) discussed the non-relativistic phase
of GRB afterglows for the first time. But at the beginning of 
the afterglow era, the Newtonian aspects of GRBs were largely 
ignored in the literature for obvious reasons. The importance 
of non-relativistic phase was stressed and then extensively 
discussed by Huang et al. (Huang et al. 1998a; Huang, Dai \& Lu 
1998b, 1999a,b; Huang et al. 2000a; Huang, Dai \& Lu 2000b,c), who
pointed out that GRB remnants enter the trans-relativistic phase 
(here we define it as $2 \leq \gamma \leq 5$) typically in a few
to 10 days. Today, the importance of Newtonian phase has been 
realized by more and more authors (e.g. Kobayashi, Piran \& 
Sari 1999; Livio \& Waxman 2000; 
Chevalier \& Li 2000; Frail, Waxman \& Kulkarni 2000; 
Weiler et al. 2002; Kulkarni 2002). 

Afterglows from both isotropic fireballs and highly collimated jets 
in the trans-relativistic and non-relativistic phases have been 
investigated in great detail by Huang et al. (1998b, 1999a,b, 2000a,b,c). 
Although their dynamical equations are valid for sufficiently
long period, their discussion cannot extend to the deep Newtonian
phase, during which most shock-accelerated electrons will cease to be 
ultra-relativistic and will no longer emit synchrotron radiation. In
this article, a more reasonable distribution function will be derived 
for electrons, so that we can extend our calculation into the deep 
Newtonian phase. The structure of our article is arranged as follows. 
We describe our dynamical and radiation model in Section 2. Based on 
the revised model, optical afterglows from isotropic fireballs, 
conical jets as well as cylindrical jets are then investigated 
numerically in Section 3. Emphases will be put on the light curve 
behaviour in the deep Newtonian phase. Section 4 is a brief discussion. 

\section{Model}

The overall dynamical evolution of GRB remnants has been studied by many
authors (Huang et al. 1999a,b; Kobayashi et al. 1999; 
Dermer \& Humi 2001; Panaitescu \& Kumar 2001a, 2002). 
Here we adopt the simple dynamical model proposed by Huang 
et al. (1999a,b), which is characterized mainly by the following 
differential equation, 
\begin{equation}
\label{eq1}
\frac{d \gamma}{d m} = - \frac{\gamma^2 - 1}
       {M_{\rm ej} + \epsilon m + 2 ( 1 - \epsilon) \gamma m}, 
\end{equation}
where $\gamma$ is the bulk Lorentz factor of the shocked interstellar 
medium (ISM), $M_{\rm ej}$ is the initial ejecta mass, $\epsilon$ is 
the radiative efficiency and $m$ is the swept-up ISM mass. It has been
shown that this equation is correct for both ultra-relativistic and 
non-relativistic blastwaves, no matter whether they are adiabatic 
or highly radiative (Huang et al. 1999a,b). In realistic GRB remnants, 
the blastwaves can be radiative only within the initial two or three 
hours following the main burst (Dai, Huang \& Lu 1999). They will become
completely adiabatic after that period. So, for simplicity, we will 
take $\epsilon \equiv 0$ directly in this study. 

For beamed ejecta with a half opening angle of $\theta$, 
Huang et al. (2000a,b) have proposed an improved method to 
describe the lateral expansion of the remnant. Their description
is based on a refined expression for the sound speed that is 
appropriate in both relativistic and non-relativistic phases. 
In Section 3, we will solve the dynamical evolution of isotropic 
fireballs, conical jets as well as cylindrical jets numerically.
For details of the numerical procedure, we refer readers to 
Huang et al. (1999b, 2000a,b) and Cheng, Huang \& Lu (2001). 

Synchrotron radiation from the shock-accelerated ISM electrons plays
a major role in the optical afterglows. To calculate this emission, 
two factors are essential to be known: the magnetic field and the 
energy of electrons. As usual, we assume that the magnetic 
energy density in the comoving frame is a fraction $\xi_{\rm B}^2$ 
of the total thermal energy density 
($B'^2 / 8 \pi = \xi_{\rm B}^2  e'$). Electrons are also usually 
assumed to carry a fraction $\xi_{\rm e}$ of the proton energy and 
follow a power-law distribution according to their Lorentz factors,
\begin{equation}
\label{eq6}
\frac{d N_{\rm e}'}{d \gamma_{\rm e}} \propto \gamma_{\rm e}^{-p} , 
\;\;\; (\gamma_{\rm e,min} \leq \gamma_{\rm e} \leq \gamma_{\rm e,max}),
\end{equation}
where the minimum Lorentz factor can be easily derived as,
\begin{equation}
\label{eq7}
\gamma_{\rm e,min} = \xi_{\rm e} \frac{m_{\rm p}}{m_{\rm e}}
\cdot \frac{p-2}{p-1} (\gamma - 1)+ 1,
\end{equation} 
with $m_{\rm p}$ and $m_{\rm e}$ being the proton and electron mass
respectively. The maximum Lorentz factor
is constrained by synchrotron cooling, 
$\gamma_{\rm e,max} \approx 10^8 (B' /1 {\rm G})^{-1/2}$ (M\'{e}sz\'{a}ros
et al. 1993; Vietri 1997; Totani 1999). Sari et al. (1998) further
suggested that radiation loss may play an important role in the 
process and can change the distribution function to 
$d N_{\rm e}' / d \gamma_{\rm e} \propto \gamma_{\rm e}^{-(p+1)}$ for 
electrons above a critical Lorentz factor $\gamma_{\rm c}$, which
is given by 
$\gamma_{\rm c} = 6 \pi m_{\rm e} c / (\sigma_{\rm T} \gamma B'^2 t)$, 
where $\sigma_{\rm T}$ is the Thomson cross section and $t$ is the time
of the observer. 
Based on these considerations, a detailed description of electron 
distribution has been presented by Dai et al. (1999, also see Huang
et al. 2000a,b).

The above discussion is valid only when electrons are ultra-relativistic,
i.e., $\gamma_{\rm e,min} \gg 1$. However, in the deep Newtonian phase,
this condition will no longer be satisfied. The bulk Lorentz factor 
($\gamma$) will be so close to unity that $\gamma_{\rm e,min}$ will also
be very close to one. It means the majority of electrons will cease to
be ultra-relativistic. For example, taking typical values of 
$\xi_{\rm e} = 0.1$ and $p = 2.2$, $\gamma_{\rm e,min}$ will equal to 
5 when $\gamma \approx 1.13$ (or $\beta = \sqrt{\gamma^2 - 1}/ \gamma
\approx 0.47$). Anyway, the 
distribution of electrons with respect to their kinetic energies can
still be assumed to be a power-law function, which now takes the 
following form, 
\begin{equation}
\label{eq8}
\frac{d N_{\rm e}'}{d \gamma_{\rm e}} \propto (\gamma_{\rm e} - 1)^{-p} , 
\;\;\; (\gamma_{\rm e,min} \leq \gamma_{\rm e} \leq \gamma_{\rm e,max}).
\end{equation}
Most electrons are now non-relativistic and their cyclotron radiation 
cannot be observed in optical bands. But there are still many 
relativistic electrons with Lorentz factor above a critical value, 
$\gamma_{\rm e,syn}$, and still capable of emitting synchrotron 
radiation. In this case, to calculate the optical afterglow, we 
just need to integrate the emission over all those electrons satisfying
$\gamma_{\rm e,syn} \leq \gamma_{\rm e} \leq \gamma_{\rm e,max}$. The
determination of $\gamma_{\rm e,syn}$ is somewhat arbitrary. We believe
that it could be some value like 5 or 10. Anyway, it is lucky that this
uncertainty will not bring any essential effect into the optical 
afterglows (see Section 3). 

Based on equation (\ref{eq8}), we can now re-derive the electron 
distribution as follows, following Dai et al.'s (1999) treatment (also
see Huang et al. 2000a,b), 
\begin{description}
\item (i) For $\gamma_{\rm c}\leq \gamma_{\rm e,min}$,
\begin{equation}
\label{dnei9}
\frac{dN_{\rm e}'}{d\gamma_{\rm e}}=C_1 (\gamma_{\rm e} - 1)^{-(p+1)}\,, \,\,\,\,\,\,
(\gamma_{\rm e,min}\leq\gamma_{\rm e}\leq \gamma_{\rm e,max})\,,
\end{equation}
\begin{equation}
\label{dneic10}
C_1=\frac{p}{(\gamma_{\rm e,min}-1)^{-p}- (\gamma_{\rm e,max}-1)^{-p}}N_{\rm ele}\,, 
\end{equation}
where $N_{\rm ele}$ is the total number of radiating electrons involved.

\item (ii) For $\gamma_{\rm e,min} < \gamma_{\rm c} \leq \gamma_{\rm e,max}$,
\begin{equation}
\label{dneii11}
  \frac{dN_{\rm e}'}{d\gamma_{\rm e}} = \left \{
   \begin{array}{ll}
 C_2 (\gamma_{\rm e}-1)^{-p}\,, \,\,\,\, & (\gamma_{\rm e,min} \leq \gamma_{\rm e}
					      \leq \gamma_{\rm c}), \\
 C_3 (\gamma_{\rm e}-1)^{-(p+1)}\,, \,\,\,\, & (\gamma_{\rm c}<\gamma_{\rm e}
					      \leq \gamma_{\rm e,max}),
   \end{array}
   \right. 
\end{equation}
where 
\begin{equation}
\label{dneiic12}
 C_2=C_3/(\gamma_{\rm c}-1)\,,
\end{equation}
$$
 C_3  =  \left[\frac{(\gamma_{\rm e,min}-1)^{1-p}-(\gamma_{\rm c}-1)^{1-p}}
                 {(p-1)(\gamma_{\rm c}-1)}   \right.
$$
\begin{equation}
\label{dneiicc13}
 \left.   + \frac{(\gamma_{\rm c}-1)^{-p}-(\gamma_{\rm e,max}-1)^{-p}}
                 {p} \right]^{-1}N_{\rm ele} .
\end{equation}

\item (iii) If $\gamma_{\rm c} > \gamma_{\rm e,max}$, then
\begin{equation}
\label{dneiii14}
\frac{dN_{\rm e}'}{d\gamma_{\rm e}}=C_4 (\gamma_{\rm e}-1)^{-p}, \,\,\,\,\,\,
(\gamma_{\rm e,min}\leq\gamma_{\rm e}\leq\gamma_{\rm e,max}),
\end{equation}
where 
\begin{equation}
\label{dneiiic15}
C_4=\frac{p-1}{(\gamma_{\rm e,min}-1)^{1-p}-(\gamma_{\rm e,max}-1)^{1-p}} N_{\rm ele}.
\end{equation}
\end{description}
Of course, $\gamma_{\rm c} \gg 1$ and $\gamma_{\rm e,max} \gg 1$ are 
safely satisfied, so that $\gamma_{\rm c} -1$ and $\gamma_{\rm e,max} -1$ 
can be simplified as $\gamma_{\rm c}$ and $\gamma_{\rm e,max}$ respectively
in these equations. 

\section{Numerical Results}

Afterglows from isotropic fireballs as well as collimated jets have been
studied in great detail by Huang et al. (Huang et al. 1999a,b, 2000a,b,c; 
Cheng, Huang \& Lu 2001). However, those discussion in general can not
extend to the deep Newtonian phase. In this section, we use our revised 
model to continue the study. In all our calculations here, we take
the following typical parameter values: 
$\gamma_{\rm e,syn} =5, \xi_{\rm e} =0.1, D_{\rm L} = 1$ Gpc and 
$p=2.2$, where $D_{\rm L}$ is the luminosity distance. The initial value 
of $\gamma$ is fixedly taken as $\gamma_0 = 300$. Other parameters, 
such as the total isotropic kinetic energy of the blastwave ($E_0$), the 
number density of the interstellar medium ($n$) and the magnetic energy 
fraction ($\xi_{\rm B}^2$), will be given separately elsewhere. 
The effects of equal arrival time surfaces (Waxman 1997c; Sari 1998; 
Panaitescu \& M\'esz\'aros 1998; Huang et al. 2000a,b) are taken into
account in our studies. For more details of the calculation, readers 
are refered to Huang et al. (1999b, 2000a,b) and Cheng, Huang \& Lu 
(2001).

\subsection{Isotropic fireballs}

Fig. 1 shows the evolution of the Lorentz factor for some exemplary 
isotropic fireballs and the corresponding $R$-band flux density
($S_{\rm R}$). Analytical solution requires that the fireball evolves
as $\gamma \propto t^{-3/8}$ in the ultra-relativistic phase and 
$\beta = \sqrt{\gamma^2 - 1}/ \gamma \propto t^{-3/5}$ 
in the Newtonian limit (Wijers et al. 1997;
Huang et al. 1999a,b). Fig. 1a consists with the requirement exactly. 
Wijers et al. (1997) and Dai \& Lu (1999, 2000) have also derived the
theoretical afterglow light curve analytically for both relativistic 
and non-relativistic phases. In case of slow cooling, we have 
\begin{equation}
\label{lc14}
  S_{\rm R} \propto \left \{
   \begin{array}{ll}
 t^{(3-3p)/4}\,, \,\,\,\, & (\gamma \gg 1), \\
 t^{(21-15p)/10}\,, \,\,\,\, & (\beta \ll 1).
   \end{array}
   \right. 
\end{equation}
Taking $p = 2.2$, we get $S_{\rm R} \propto t^{-0.9}$ and 
$S_{\rm R} \propto t^{-1.2}$ for relativistic and Newtonian phases 
respectively. Our numerical results are consistent with their solutions. 
For example, the slope of the solid line in Fig. 1b is $\sim -0.92$ and
$\sim -1.22$ in the relativistic phase and the non-relativistic phase 
respectively. 

Fig. 1 also clearly shows that the remnant enters the deep Newtonian phase
in a relatively short period. For example, $\gamma_{\rm e,min}$ becomes
less than $\gamma_{\rm e,syn}$ at $\sim 10^7$ s, i.e., about 3 months. 
Since optical afterglows from some GRBs have been detected for more than
six months, and radio afterglows are even detectable one or three years 
later (Frail et al. 2000; Kulkarni 2002), we see that the 
study of afterglows in the deep Newtonian phase is really essential. 

In our calculations, we have assumed that the minimum Lorentz factor of 
electrons capable of emitting synchrotron radiation is 
$\gamma_{\rm e,syn} = 5$. This evaluation is somewhat arbitrary, but it 
does not affect the optical light curve too much. The characteristic 
synchrotron frequency of an electron with Lorentz factor $\gamma_{\rm e}$
in a magnetic field $B'$ is 
$\nu = \gamma_{\rm e}^2 e B'/ (2 \pi m_{\rm e} c) \approx 
2.8 \times 10^6 \gamma_{\rm e}^2 (B'/1 {\rm G}) $ Hz. Electrons in the 
lower energy range thus do not contribute to optical flux density 
significantly. In fact, we have taken $\gamma_{\rm e,syn}$ as 10 or 
even 50 in our trial calculations and just found no difference in the 
optical light curves.

\subsection{Conical jets}

Collimation is very important in GRBs, which can affect the intrinsic
kinetic energy and may provide direct clues on the progenitors 
(Frail et al. 2001; Ioka \& Nakamura 2001; 
Panaitescu \& Kumar 2001a,b, 2002; Ramirez-Ruiz 
\& Lloyd-Ronning 2002; Yamazaki, Ioka \& Nakamura 2002). Beaming 
effects in afterglows have been discussed extensively in the 
literature. It is generally believed that due to both the edge 
effect (Kulkarni et al. 1999; M\'esz\'aros \& Rees 1999; Panaitescu 
\& M\'esz\'aros 1999) and the lateral expansion effect (Rhoads 1997, 
1999), afterglows from a conical jet are characterized by a break 
in the light curve. The break point is approximately determined by
$\gamma \sim 1/\theta_0$, where $\theta_0$ is the initial half 
opening angle. GRBs 990123, 990510 and 000301c are 
regarded as good examples (Castro-Tirado et al. 1999; Harrison et 
al. 1999; Kulkarni et al. 1999; Sari, Piran \& Halpern 1999; Wijers
et al. 1999; Masetti et al. 2000; Holland et al. 2000; 
Berger et al. 2000; Jensen et al. 2001; Huang et al. 2000c). However, 
detailed numerical studies show that the break is usually quite 
smooth (Moderski, Sikora \& Bulik
2000). Huang et al. further found that the break is 
parameter-dependent (Huang et al. 2000a,b). 
They also suggested that the trans-relativistic and non-relativistic
stages are important and should be considered carefully. In this 
section, we use our revised model to reveal the behaviour of conical
jets in the deep Newtonian phase. For more details of the calculation, 
please refer to Huang et al. (2000a,b). 

\subsubsection{Conical jets without lateral expansion}

Fig. 2 illustrates the evolution and afterglows of some exemplary 
conical jets without lateral expansion. In fact, the dynamical evolution
of such a jet should be very similar to that of an isotropic fireball, 
i.e., $\gamma \propto t^{-3/8}$ in the ultra-relativistic phase and 
$\beta \propto t^{-3/5}$ in the Newtonian stage. Fig. 2a shows these
trends exactly. As for the optical light curve, analytical solutions 
predict that it should follow $S_{\rm R} \propto t^{(3-3p)/4}$ 
($t^{-0.9}$ for $p=2.2$) before $\gamma = 1/\theta_0 =10$. But soon 
after the $\gamma = 10$ point, a break should appear and the light curve 
becomes $S_{\rm R} \propto t^{-3p/4}$ ($t^{-1.65}$ for $p = 2.2$). 
This relation can be derived as follows. 
Analytically we have $\gamma \propto t^{-3/8}$ and $R \propto t^{1/4}$. 
Then the characteristic synchrotron frequency is 
$\nu_{\rm m} \propto \gamma \gamma_{\rm e,min}^2 B' \propto \gamma^4 
\propto t^{-3/2}$, and the corresponding maximum flux density is
$S_{\nu,{\rm max}} \propto N_{\rm ele} \gamma B' / \gamma^{-2} \propto 
R^3 \gamma^4 \propto t^{-3/4}$, where the $\gamma^{-2}$ item in 
$S_{\nu,{\rm max}}$ arises from the edge effect when $\gamma < 
1 / \theta_0$. So, the observed optical
flux density is $S_{\rm R} \approx S_{\nu,{\rm max}} (\nu_{\rm R} / 
\nu_{\rm m})^{-(p-1)/2} \propto t^{-3p/4}$. 
Such a light curve break can be clearly seen in Fig. 2b. 
For example, the solid line can be fit as 
$S_{\rm R} \propto t^{-0.89}$ during $10^3$ --- $10^4$ s and it is 
approximately $S_{\rm R} \propto t^{-1.80}$ during $10^5$ --- $10^6$ s.
The slopes are acceptably consistent with analytic values. 
However, it should be noted that the break occurs obviously 
later than the time determined by $\gamma = 1/\theta_0 =10$, which has 
been pointed out by Huang et al. in previous studies 
(Huang et al. 2000a,b). 

Maybe the most striking feature in Fig. 2b is the flattening of the 
light curves in the deep Newtonian phase. This is not difficult to 
understand. In such a non-relativistic phase, the remnant, although 
beamed, should behaves more or less like a spherical shell since the 
edge effect no longer exists (Livio \& Waxman 2000). Beaming will only
reduce the total flux density by a constant factor. In fact, the slope
of the solid line is approximately $-1.2$ after $10^9$ s, which is 
just the value expected for an isotropic fireball (see Section 3.1). 
So, we see that the overall evolution of optical afterglows from a 
conical jet without lateral expansion can be typically described by, 
\begin{equation}
\label{eqlc17}
  S_{\rm R} \propto \left \{
   \begin{array}{ll}
 t^{(3-3p)/4}\,, \,\,\,\, & (\gamma > 1/ \theta_0), \\
 t^{-3p/4}\,, \,\,\,\, & (\gamma < 1/\theta_0 \; {\rm and} \; \beta \sim 1), \\
 t^{(21-15p)/10}\,, \,\,\,\, & (\beta \ll 1).
   \end{array}
   \right.
\end{equation}

\subsubsection{Conical jets with lateral expansion}

In realistic cases, jets may expand laterally at comoving sound speed. 
Fig. 3a shows the evolution of Lorentz factor for such realistic 
conical jets. The curves can be approximately fit by $\gamma
\propto t^{-0.39}$ in the highly relativistic phase and by 
$\gamma -1 \propto t^{-1.25}$ in the non-relativistic phase, 
consistent with the analytic results of $\gamma \propto t^{-3/4}$
and $\gamma - 1 \propto t^{-6/5}$ respectively. Again we see that
$\gamma_{\rm e,min}$ becomes less than 5 when $t \geq 3 \times 10^6$ 
--- $2 \times 10^7$ s. 

Fig. 3b illustrates the $R$-band afterglows correspondingly. We see
that the general behaviour of the light curves is very similar to that
in Fig. 2b. However, slight difference can still be found and deserves 
addressing in some detail. For example, the solid line approximately 
follows $S_{\rm R} \propto t^{-1.03}$ in the ultra-relativistic phase, 
which is steeper by $\sim t^{0.14}$ than that in Fig. 2b. It indicates 
that the lateral expansion tends to make the afterglow decay faster. 
Subsequently the light curve steepens to $S_{\rm R} \propto t^{-2.24}$,
which is also steeper by $\sim t^{0.44}$ than that in Fig. 2b.  
It is clear that lateral expansion contributes to the light curve break 
significantly.
In the deep Newtonian phase, the light curves are again markedly 
characterized by a flattening. The slope of the solid line in this segment
is $\sim -1.26$, also consistent with the analytical result of $-1.2$ for
an isotropic fireball. In short, we conclude that the overall evolution 
of optical afterglows from a conical jet with lateral expansion can be 
typically expressed as,
\begin{equation}
\label{eqlc17a}
  S_{\rm R} \propto \left \{
   \begin{array}{ll}
 t^{(3-3p)/4}\,, \,\,\,\, & (\gamma > 1/ \theta_0), \\
 t^{-p}\,, \,\,\,\, & (\gamma < 1/\theta_0 \; {\rm and} \; \beta \sim 1), \\
 t^{(21-15p)/10}\,, \,\,\,\, & (\beta \ll 1).
   \end{array}
   \right.
\end{equation}

\subsection{Cylindrical jets}

The geometry of GRB jets is usually assumed to be conical. However, many
of the relativistic outflows in radio galaxies are found to maintain 
constant cross-sections at large scales (Perley, Bridle \& Willis 1984;
Biretta, Sparks \& Macchetto 1999). This has led Cheng et al.  
(2001) to suggest that GRB jets might also be cylindrical. In fact, 
a very similar idea, i.e., the so called cannon-ball model, has been 
suggested as the GRB trigger mechanism by Dar et al. (Shaviv \& Dar 
1995; Dar 1998), and observed afterglows have also been examined 
carefully in this frame (Dado, Dar \& De R\'ujula 2002). 

Theoretical GRB afterglows from cylindrical jets 
have been studied by Cheng et al.  
(2001) in great detail. However, their discussion again cannot 
extend to the deep Newtonian phase. Here we use our revised model to 
continue the study. Since a cylindrical jet without lateral expansion
generally decelerates very slowly and can remain to be highly 
relativistic for as long as $\sim 10^8$ --- $10^9$ s (Cheng et al. 
2001), we will discuss only cylindrical jets with lateral expansion. 

Fig. 4a shows the evolution of $\gamma$ for some exemplary cylindrical
jets. In the highly relativistic phase the curves approximately follow
$\gamma \propto t^{-0.5}$, and in the non-relativistic phase they follow
$\gamma -1 \propto t ^{-1.17}$. The timing indices are consistent with 
the analytical results of $-1/2$ and $-6/5$ respectively (Cheng et al.  
2001). Note that in this figure, $\gamma_{\rm e,min}$ is already 
less than 5 after $\sim $ (2 --- 4) $\times 10^6$ s. Fig. 4b illustrates
the evolution of optical afterglows correspondingly. In the 
ultra-relativistic phase, the solid line can be approximated as 
$S_{\rm R} \propto t^{-2.34}$, consisting with the analytical solution of
$S_{\rm R} \propto t^{-p}$ (Cheng et al. 2001). After entering the
non-relativistic phase, the remnant is expected to resemble an isotropic
fireball, so that the light curve should be $S_{\rm R} \propto 
t^{(21-15p)/10}$ (Cheng et al. 2001). Taking $p=2.2$, we get the 
analytical timing index of $-1.2$. In fact, such a flattening is really 
observed in Fig. 4b, and the slope of the solid line is just $\sim -1.33$
in the deep Newtonian phase. From these studies, we see that the optical 
afterglow from laterally expanding cylindrical jets can typically be well 
represented by,  
\begin{equation}
\label{eqlc18}
  S_{\rm R} \propto \left \{
   \begin{array}{ll}
 t^{-p}\,, \,\,\,\, & (\gamma \gg 1), \\
 t^{(21-15p)/10}\,, \,\,\,\, & (\beta \ll 1).
   \end{array}
   \right. 
\end{equation}

\section{Discussion and Conclusions}

We have shown that typically in $\sim 10^7$ s, the majority of 
shock-accelerated electrons will become non-relativistic, i.e., 
$\gamma_{\rm e,min} \leq \gamma_{\rm e,syn}$. These electrons can only 
emit cyclotron radiation which falls mainly in the frequency range of 
$\sim 10^6$ --- $10^7$ Hz. However, a small portion of electrons are 
still highly relativistic and capable of emitting synchrotron radiation. 
A revised distribution function has been derived for electrons, which
allows us to calculate the emission from these relativistic particles. 

Optical afterglows from both isotropic fireballs and highly collimated jets 
are studied numerically based on the revised model. For isotropic fireballs,
the optical light curve steepens by $t^{(15p-27)/20}$ after entering the 
deep Newtonian phase. This 
corresponds to $\sim t^{0.3}$ for $p=2.2$, or $\sim t^{0.9}$ for $p=3$. 
For conical jets, we find that the light curve steepens slightly later than
the moment determined by $\gamma = 1/\theta_0$. It is clear that this 
steepening is due to the edge effect and is strengthened by lateral 
expansion. Additionally, the optical light 
curves are universally characterized by a striking flattening during the 
deep Newtonian phase, which can be attributed to the resembling of the 
remnant to an isotropic fireball at such late stages (Livio \& Waxman
2000). The flattening is also present in the light curves of cylindrical 
jets. 

Long-lasting optical afterglows have been observed from a number of GRBs, 
e.g. GRBs 970228, 970508, 991208 (Fruchter et al. 1999, 2000;
Castro-Tirado et al. 2001). Optical 
transients in these events are typically detectable for $\sim 100$ --- 
200 d. Our studies here are of obvious importance in such cases. 
However, the flattening of the light curve in the deep Newtonian phase 
predicted for highly collimated jets has not been observed in realistic 
observations of jet candidates. A possible reason may be that the optical 
flux densities are usually too faint. In fact, to observe this flattening
clearly, we should follow the optical transient for at least $\sim 10^8$ 
s, i.e., more than $\sim 3$ years. This could hardly be actualized even with
the Keck telescope and the Hubble Space Telescope. Additionally, optical 
emission from the host galaxy makes the measurement even more difficult. 
Anyway, the next generation optical telescopes may cast some 
light on the problem. 

Radio afterglows can hopefully be observed for even longer period. In fact, 
a 450-day light curve of the radio afterglow of GRB 970508 has been reported 
by Frail et al. (2000). GRB 980703 has even been monitored in
radio bands for more than 1000 days (Kulkarni 2002). At such stages, the 
GRB remnant should have entered the deep Newtonian phase. Observations of 
radio afterglows on such a late stage have the priority that they allow for
a direct measure of the intrinsic kinetic energy (Frail et al. 
2000). The effect of $\gamma_{\rm e,min}$ will have to be considered in 
fitting those radio light curves theoretically. Calculation of radio emission
involves synchrotron self absorption, which is beyond the scope of this 
article and will be discussed elsewhere. 

Orphan afterglow survey is thought to be a useful method to determine the 
degree of beaming in GRBs (Rhoads 1997; M\'esz\'aros, Rees \& Wijers 1999; 
Totani \& Panaitescu 2002; Nakar, Piran \& Granot 2002; Granot et al. 2002). 
However, the method is also troubled by many problems (Dalal, Griest \& 
Pruet 2002; Huang, Dai \& Lu 2002; Levinson et al. 2002; Nakar \& Piran 2003). 
For example, as suggested by Huang et al. (2002), the method is greatly
complicated by the possibility that there might be many failed GRBs, i.e.,
isotropic fireballs with $ 1 \ll \gamma \ll 100$, which cannot successfully 
generate $\gamma$-ray bursts but do can produce afterglows. Anyway, a 
successful application of the method would involve two important ingredients:
the orphan needs to be discovered as early as possible and needs to be 
followed as long as possible. In fact, it has been suggested that radio 
searches, which are capable of finding potential GRB remnants as old as 
hundreds of days, may provide the best hope to find the missing orphans 
(Dalal et al. 2002; Totani \& Panaitescu 2002; Levinson et al. 
2002). Again we see that theoretical investigation of GRB afterglows in 
the deep Newtonian phase is essential. 

\section*{Acknowledgments}
We thank an anonymous referee for valuable comments and suggestions.
YFH also thanks Y. Lu and J. C. S. Pun for stimulating discussion. 
This research was supported by a RGC grant of Hong Kong SAR, 
the National Natural Science Foundation of China, the Special 
Funds for Major State Basic Research Projects, and the Foundation 
for the Author of National Excellent Doctoral Dissertation of 
P. R. China (Project No: 200125).


{}


\begin{figure} \centering 
\epsfig{file=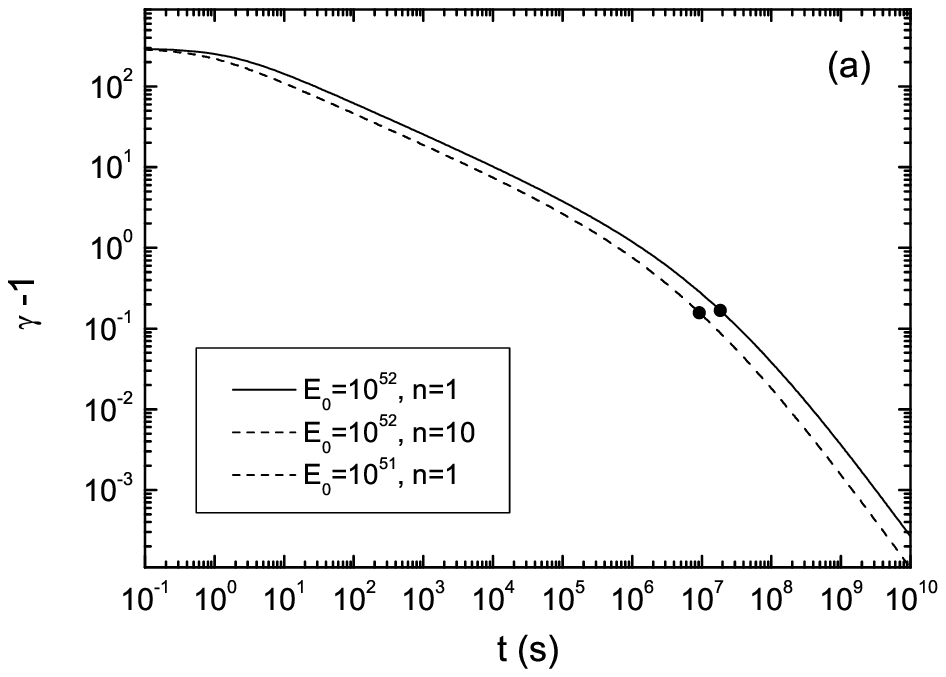, angle=0, height=70mm, width=6.8cm, 
bbllx=40pt, bblly=10pt, bburx=260pt, bbury=220pt} \\
\epsfig{file=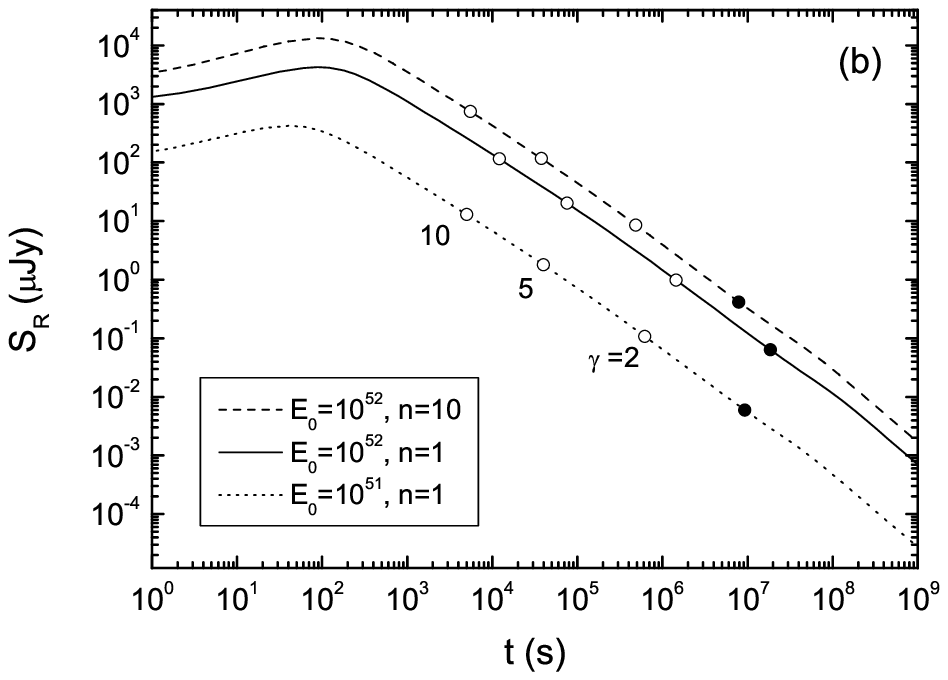, angle=0, height=70mm, width=6.8cm, 
bbllx=40pt, bblly=10pt, bburx=260pt, bbury=220pt}
\caption{ Evolution of the Lorentz factor for isotropic fireballs (a) 
  and the corresponding $R$-band optical afterglows (b). We take 
  $\xi_{\rm B}^2 = 10^{-4}$. $E_0$ and $n$ values are marked in units
  of erg and cm$^{-3}$ respectively. Other parameters are evaluated 
  in the first paragraph of Section 3. The black dot on each curve
  indicates the moment when $\gamma_{\rm e,min} = \gamma_{\rm e,syn}$, 
  and the open circles on the light curves mark the time when 
  $\gamma =2, 5$ and 10 respectively. Note that the evolution of 
  $\gamma$ is almost identical for the cases of $E_0 = 10^{52}$ erg,
  $n = 10$ cm$^{-3}$ and $E_0 = 10^{51}$ erg, $n = 1$ cm$^{-3}$.  }
\label{fig1}
\end{figure}

\begin{figure} \centering
\epsfig{file=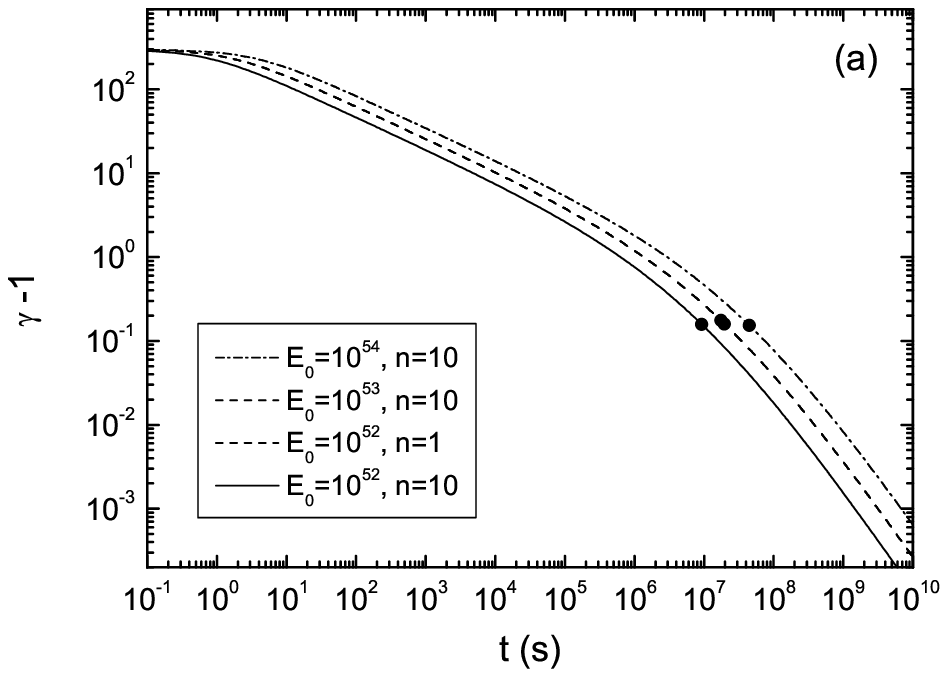, angle=0, height=70mm, width=6.8cm, 
bbllx=40pt, bblly=10pt, bburx=260pt, bbury=220pt} \\
\epsfig{file=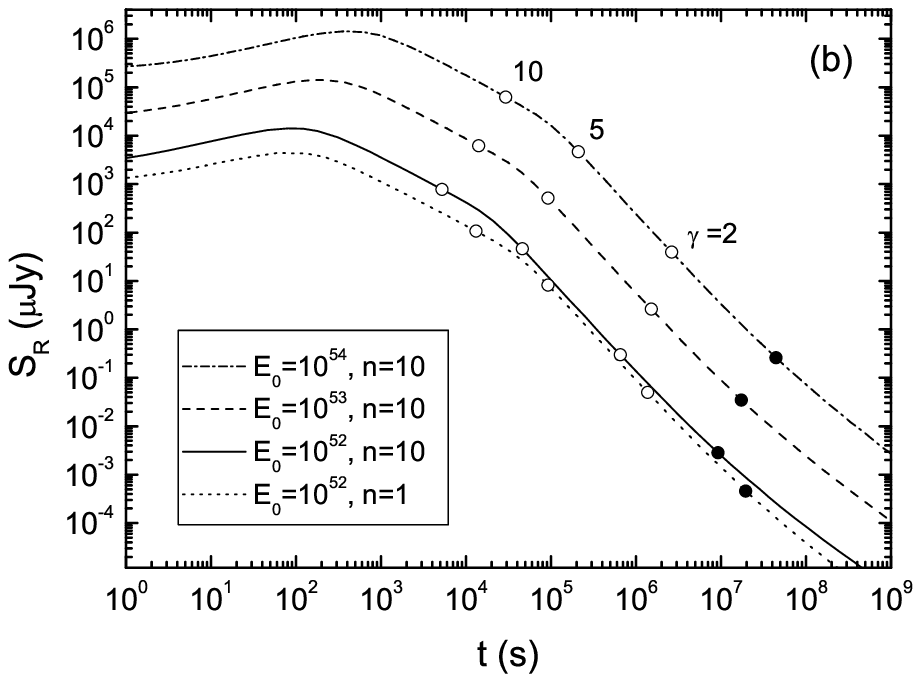, angle=0, height=70mm, width=6.8cm, 
bbllx=40pt, bblly=10pt, bburx=260pt, bbury=220pt}
\caption{ Evolution of the Lorentz factor for conical jets without 
  lateral expansion (a) and the corresponding optical afterglows (b). 
  We take $\theta_0 = 0.1, \xi_{\rm B}^2 = 10^{-4}$. Isotropic 
  equivalent energy $E_0$ and $n$ values are marked in units of erg 
  and cm$^{-3}$ respectively. Other parameters are evaluated in the 
  first paragraph of Section 3. Observers are assumed to be on the 
  axis of the jet. The black dots and the open circles have the 
  same meaning as in Fig. 1. Note that the evolution of $\gamma$ is
  almost identical for the cases of $E_0 = 10^{53}$ erg, $n=10$ 
  cm$^{-3}$ and $E_0 = 10^{52}$ erg, $n = 1$ cm$^{-3}$.}
\label{fig2}
\end{figure}

\begin{figure} \centering 
\epsfig{file=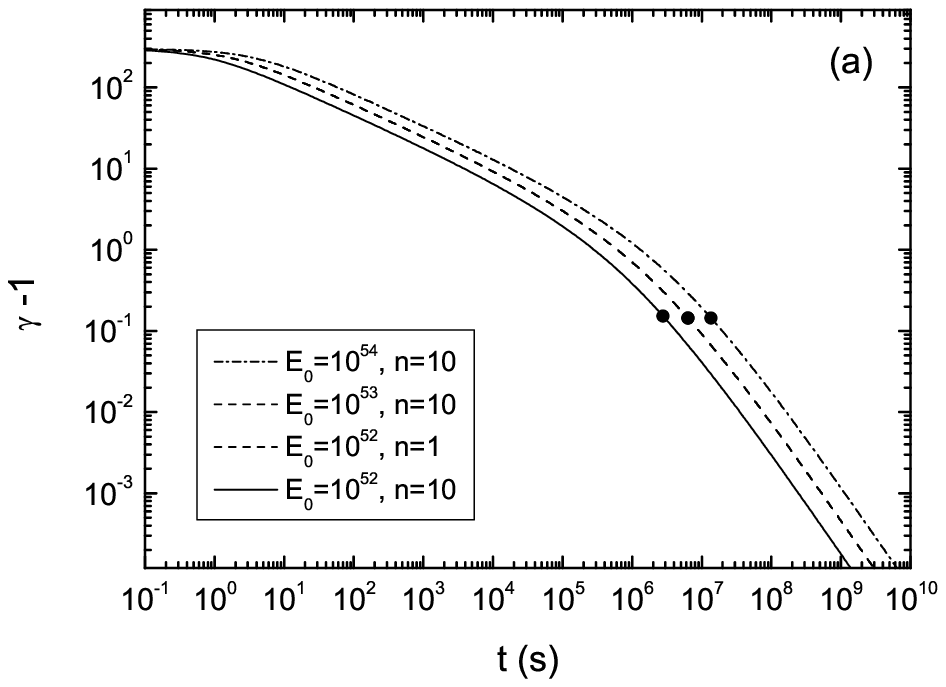, angle=0, height=70mm, width=6.8cm, 
bbllx=40pt, bblly=10pt, bburx=260pt, bbury=220pt}  \\
\epsfig{file=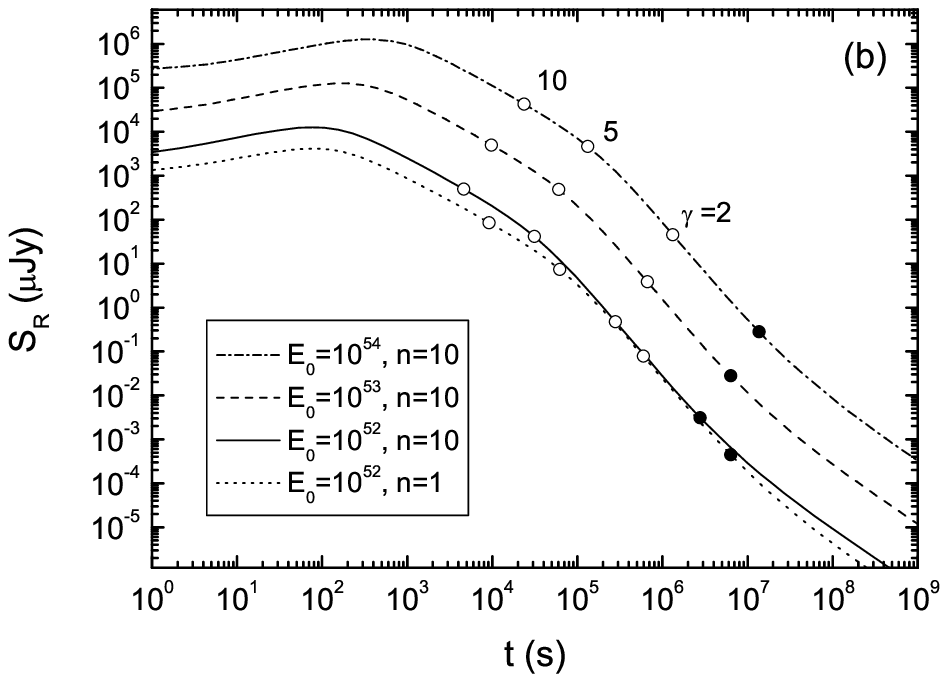, angle=0, height=70mm, width=6.8cm, 
bbllx=40pt, bblly=10pt, bburx=260pt, bbury=220pt}
\caption{ Same as Fig. 2 except that the conical jets here are 
  expanding laterally at comoving sound speed. }
\label{fig3}
\end{figure}

\begin{figure} \centering
\epsfig{file=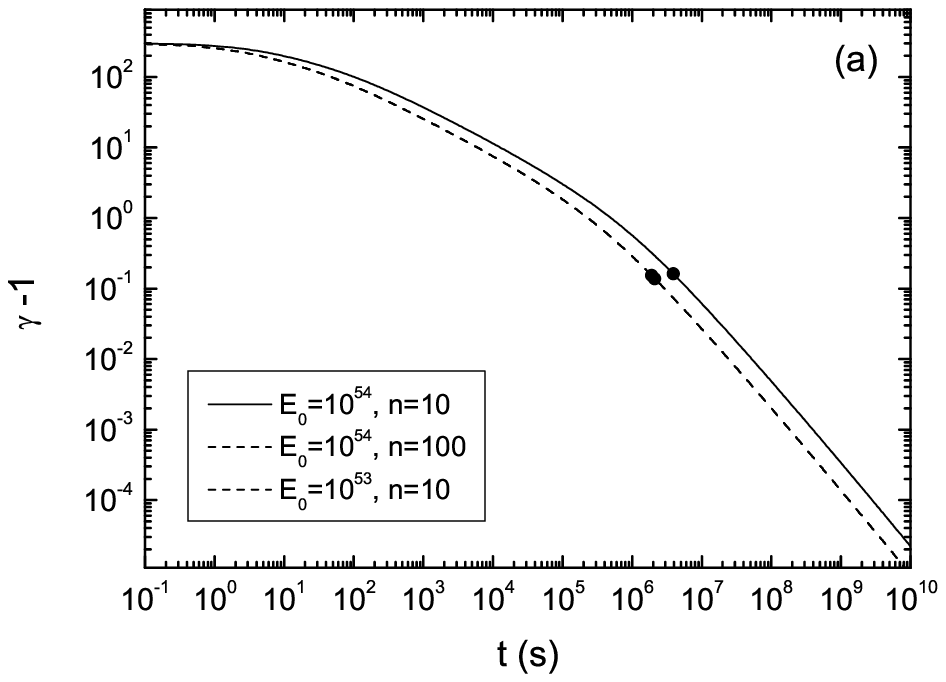, angle=0, height=70mm, width=6.8cm, 
bbllx=40pt, bblly=10pt, bburx=260pt, bbury=220pt} \\
\epsfig{file=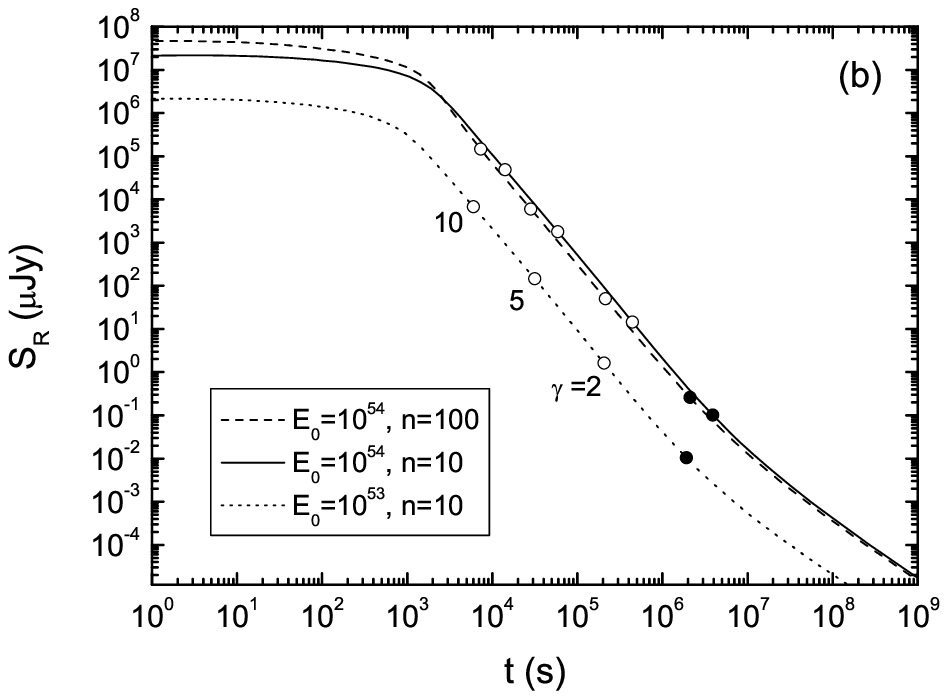, angle=0, height=70mm, width=6.8cm, 
bbllx=40pt, bblly=10pt, bburx=260pt, bbury=220pt}
\caption{ Evolution of the Lorentz factor for cylindrical jets with 
  lateral expansion (a) and the corresponding optical afterglows (b). 
  We take $\xi_{\rm B}^2 = 0.01$ and assume that the initial comoving 
  lateral radius of the jet is $a_0 = 0.01 R_0$. Isotropic
  equivalent energy $E_0$ and number density $n$ are marked in units of
  erg and cm$^{-3}$ respectively. Other parameters are evaluated in the
  first paragraph of Section 3. Observers are assumed to be on the axis
  of the jet. The black dots and the open circles have the same 
  meaning as in Fig. 1. Note that the evolution of $\gamma$ is
  almost identical for the cases of $E_0 = 10^{54}$ erg, $n=100$ 
  cm$^{-3}$ and $E_0 = 10^{53}$ erg, $n = 10$ cm$^{-3}$. }
\label{fig4}
\end{figure}

\end{document}